\documentclass[twoside,slac_one]{revtex4}
\usepackage{graphicx}
\usepackage{fancyhdr}
\usepackage{amsmath} % American Mathematics Society standards
\usepackage{bm}% bold math
\usepackage{amsxtra}
\usepackage{amssymb}
\usepackage{amsthm}
\usepackage{latexsym}
\usepackage{lscape}

\newcommand{\q}[2]{\ensuremath{#1\ \mathrm{#2}}}

\begin{document}

%Title of paper
\title{New Methods of Particle Collimation in Colliders}

% Repeat the \author .. \affiliation  etc. as needed
%
% \affiliation command applies to all authors since the last
% \affiliation command. The \affiliation command should follow the
% other information

\author{Giulio Stancari\footnote{On leave from Istituto Nazionale di
    Fisica Nucleare (INFN), Sezione di Ferrara, Italy.}}

\affiliation{Fermi National Accelerator Laboratory, P.O. Box 500,
  Batavia, Illinois 60510, USA}

\begin{abstract}
  The collimation system is an essential part of the design of any
  high-power accelerator. Its functions include protection of
  components from accidental and intentional energy deposition,
  reduction of backgrounds, and beam diagnostics. Conventional
  multi-stage systems based on scatterers and absorbers offer robust
  shielding and efficient collection of losses. Two complementary
  concepts have been proposed to address some of the limitations of
  conventional systems: channeling and volume reflection in bent
  crystals and collimation with hollow electron beams.  The main focus
  of this paper is the hollow electron beam collimator, a novel
  concept based on the interaction of the circulating beam with a
  5-keV, magnetically confined, pulsed hollow electron beam in a
  2-m-long section of the ring. The electrons enclose the circulating
  beam, kicking halo particles transversely and leaving the beam core
  unperturbed. By acting as a tunable diffusion enhancer and not as a
  hard aperture limitation, the hollow electron beam collimator
  extends conventional collimation systems beyond the intensity limits
  imposed by tolerable losses. The concept was tested experimentally
  at the Fermilab Tevatron proton-antiproton collider. Results on the
  collimation of 980-GeV antiprotons are presented, together with
  prospects for the future.
\end{abstract}

%\maketitle must follow title, authors, abstract
\maketitle

\thispagestyle{fancy}

% body of paper here - Use proper section commands
% References should be done using the \cite, \ref, and \label commands
% Put \label in argument of \section for cross-referencing
%\section{\label{}}

%%%%%%%%%%%%%%%%%%%%%%%%%%%%%%%%%%

\section{Introduction}

In high-power accelerators, the stored beam energy can be large: about
2~MJ in the Tevatron, and several hundred megajoules in the LHC at
nominal energies and intensities. Uncontrolled losses of even a small
fraction of particles can damage components, cause magnets to lose
superconductivity, and increase experimental backgrounds. Contributing
to these losses is the beam halo, continuously replenished by beam-gas
and intrabeam scattering, ground motion, electrical noise in the
accelerating cavities, resonances and, in the case of colliders,
beam-beam forces.

The beam collimation system is therefore vital for the operation of
high-power machines. Conventional collimation schemes include
scatterers and absorbers, possibly including several stages. Primary
collimators are the devices closest to the beam. They impart random
transverse kicks, mainly due to multiple Coulomb scattering, to
particles in the halo. The affected particles have increasing
oscillation amplitudes and a large fraction of them is captured by the
secondary collimators. These systems offer robust shielding of
sensitive components. They are also very efficient in reducing beam
losses at the experiments. As an example, a description of the
Tevatron collimation system can be found in
Ref.~\cite{Mokhov:JINST:2011}.

The classic multi-stage system does have limitiations: a fraction of
particles is always lost around the ring (leakage); collimator jaws
have an electromagnetic impedance (wakefields); and high losses are
generated during collimator setup when the jaws are moved
inward. Another problem is beam jitter. The orbit of the circulating
beam oscillates due ground motion and other vibrations. Even with
active orbit stabilization, the beam centroid can oscillate by tens of
microns. This translates into periodic bursts of losses at aperture
restrictions.

Bent crystals and the hollow electron beam collimator are two advanced
techniques to address these limitations in complementary
ways. Channeling and volume reflection in bent crystals reduce leakage
by directing halo particles deeper into the absorbers in a single
pass~\cite{Tsyganov:FNAL:1976, Maslov:SSCL:1991, Scandale:PRL:2009,
  Zvoda:PAC:2011, Scandale:IPAC:2011}. They replace the random kicks
of multiple scattering with well-defined deflections.

In this paper, we focus on the nature of the hollow electron beam
collimator and on the experiments conducted in the Tevatron collider
to study whether it is a viable complement to conventional systems.

\section{The hollow electron beam collimator}

\begin{figure}[bpt]
\includegraphics[width=0.56\textwidth]{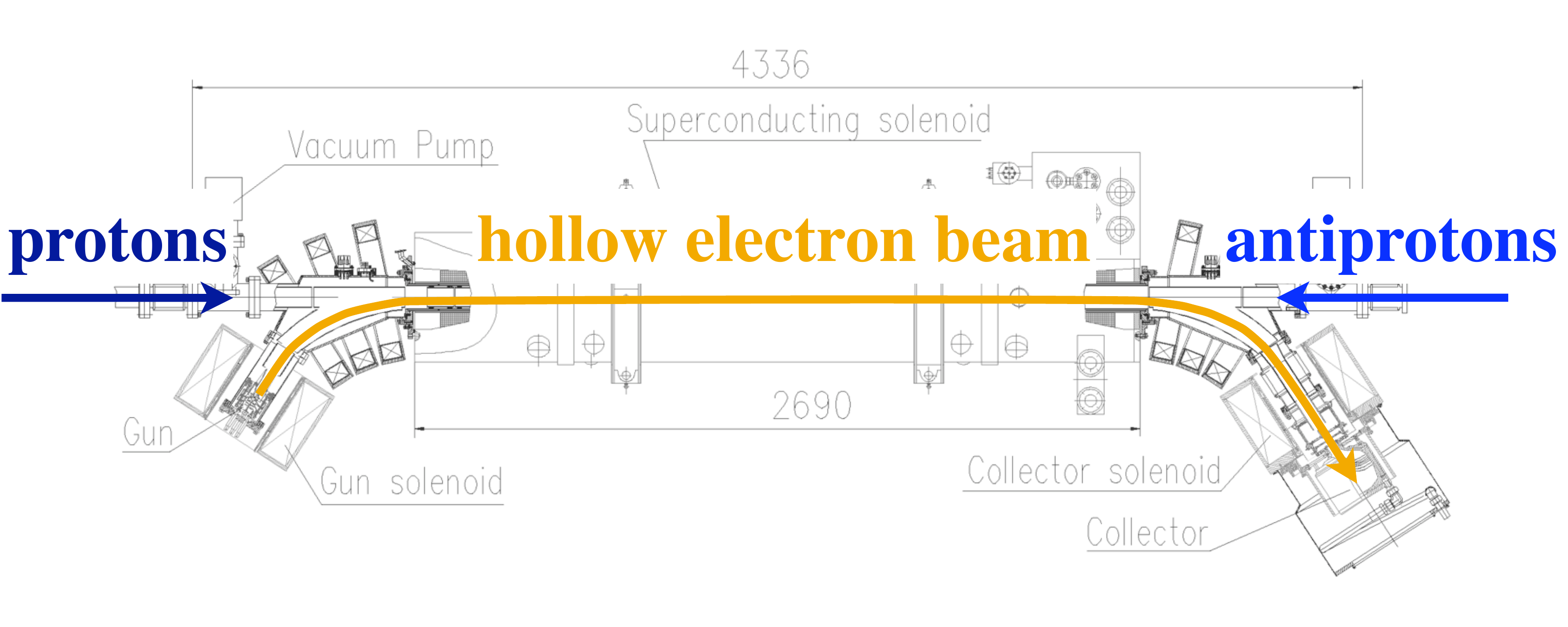} \hfill
\includegraphics[width=0.40\textwidth]{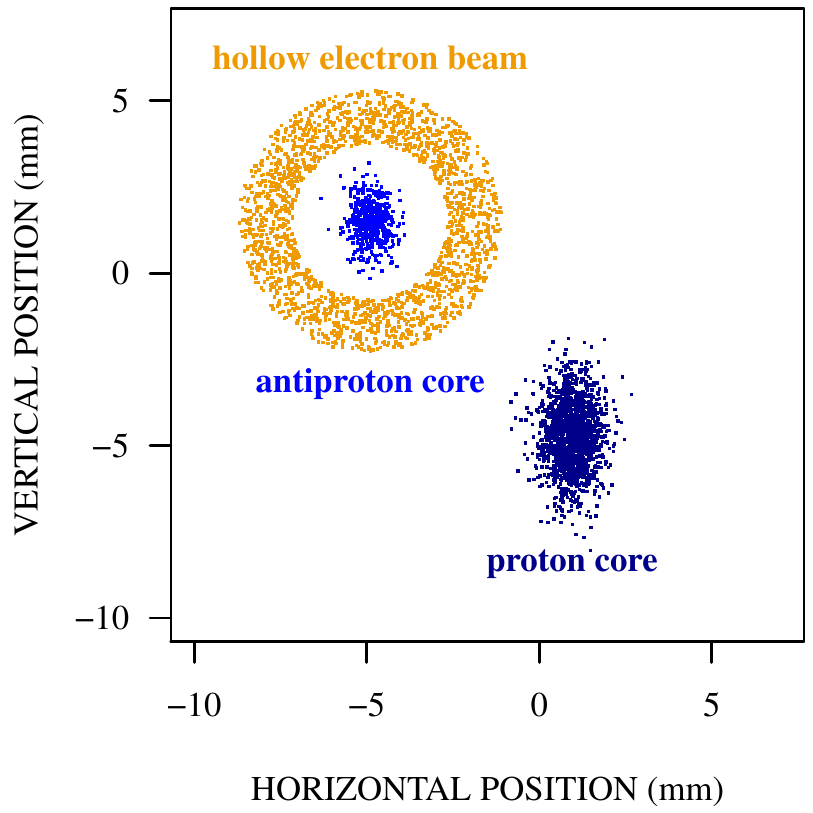}
\caption{Hollow electron beam collimator: (left)~top view;
  (right)~schematic transverse view of the three
  beams.\label{fig:hebc.concept}}
\end{figure}

\begin{figure}[bpt]
\includegraphics[width=0.48\textwidth]{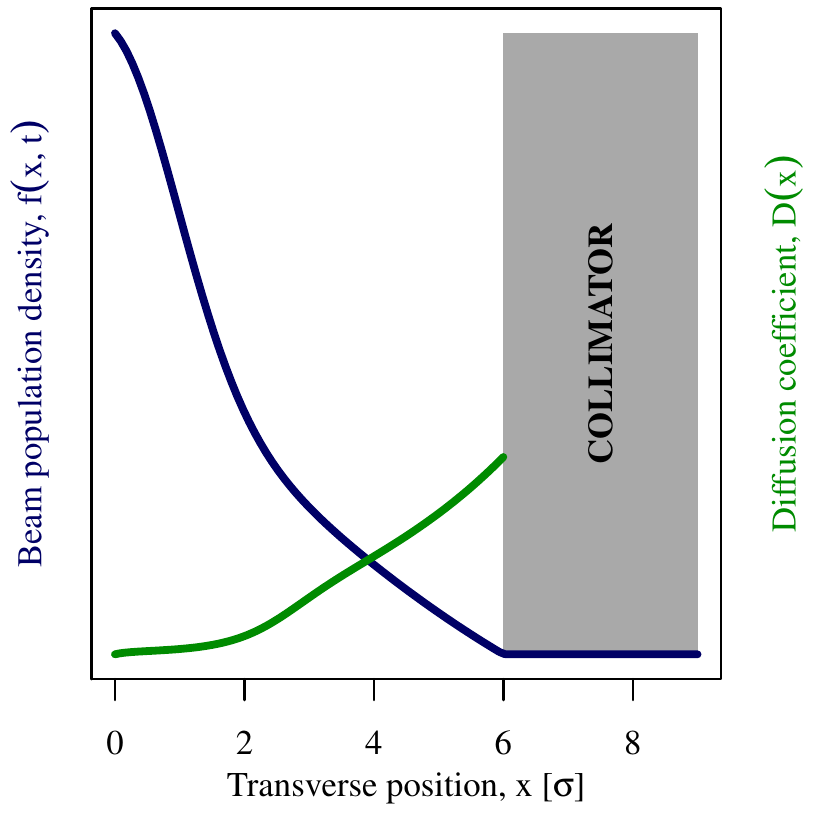}
\hfill
\includegraphics[width=0.48\textwidth]{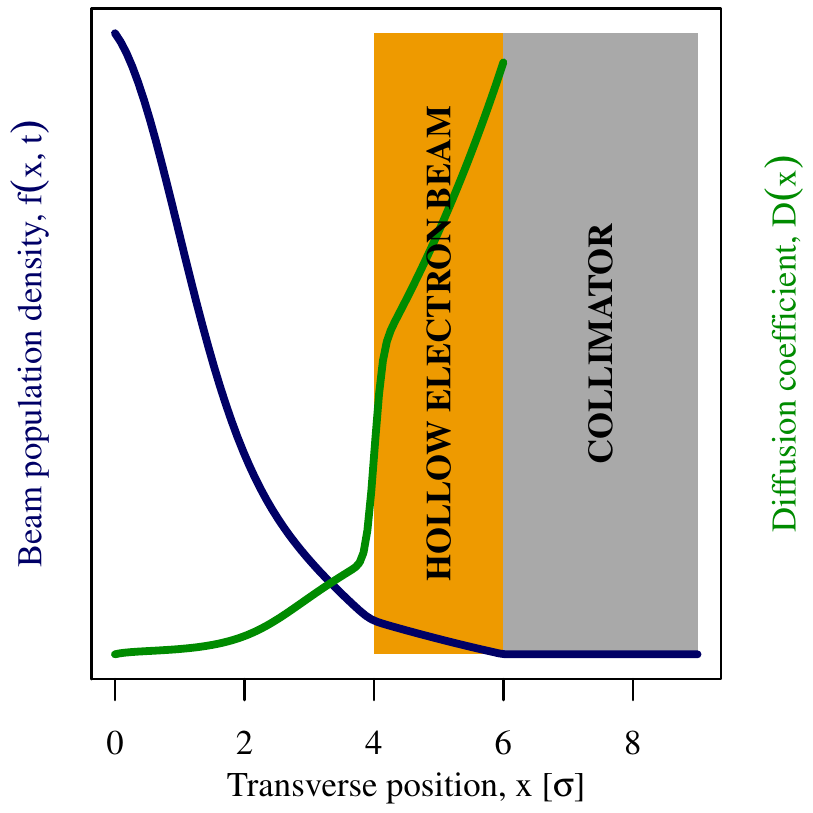}
\caption{Schematic representation of the diffusion model of
  collimation with and without the hollow electron
  beam.\label{fig:diffusion}}
\end{figure}

The hollow electron beam collimator (HEBC) is a cylindrical, hollow,
magnetically confined, possibly pulsed electron beam overlapping with
the beam halo~\cite{Shiltsev:CERN:2007a, Shiltsev:CERN:2007b,
  Shiltsev:EPAC:2008, Smith:PAC:2009, Stancari:IPAC:2010,
  Stancari:AIP:2010} (Figure~\ref{fig:hebc.concept}). Electrons
enclose the circulating beam. Halo particles are kicked transversely
by the electromagnetic field of the electrons. If the hollow charge
distribution is axially symmetric, the core of the circulating beam
does not experience any electric or magnetic fields. For typical
parameters, the transverse kick given to 980-GeV protons or
antiprotons is of the order of \q{0.2}{\mu rad}. This is to be
compared with the multiple-scattering random kick of \q{17}{\mu rad}
from the primary tungsten collimators in the Tevatron.

The effect of the electron lens can be understood in terms of a
diffusion model of collimation (Figure~\ref{fig:diffusion}). The beam
distribution has a core with long tails (the halo). The diffusion
coefficient is an increasing function of amplitude because of lattice
and beam-beam nonlinearities. The collimators define the aperture of
the machine, or the point where the population density is practically
zero. In the diffusion model, the local loss rate is the particle flux
at the collimator. It is proportional to the product of diffusion rate
and density gradient. With the hollow electron lens, one aims at
enhancing diffusion of the tails, reducing their population. This will
decrease the loss spikes caused by collimator setup and by beam
jitter.

A magnetically confined electron beam has several advantages. It can
be placed very close to, and even overlap with the circulating
beam. The transverse kicks are small and tunable, so that the device
acts more like a ``soft collimator'' or a ``diffusion enhancer,''
rather than a hard aperture limitation.  At even higher electron
currents (which have not been demonstrated so far) the electron beam
could become an indestructible primary collimator. If needed, the
electron beam can be pulsed resonantly with the betatron oscillations
to remove particles faster. In the case of ion collimation, there is
no nuclear breakup. Finally, the device relies on the estabilished
technologies of electron cooling~\cite{Parkhomchuk:RAST:2008} and
electron lenses~\cite{Shiltsev:PRL:2007, Shiltsev:NJP:2008,
  Shiltsev:PRSTAB:2008, Zhang:PRSTAB:2008}. One disadvantage may be
the cost and complexity of the required components, such as
superconducting solenoids, magnet and high-voltage power supplies, and
cryogenics.

\section{Beam experiments at the Tevatron}

\begin{figure}[b]
\includegraphics[width=0.48\textwidth]{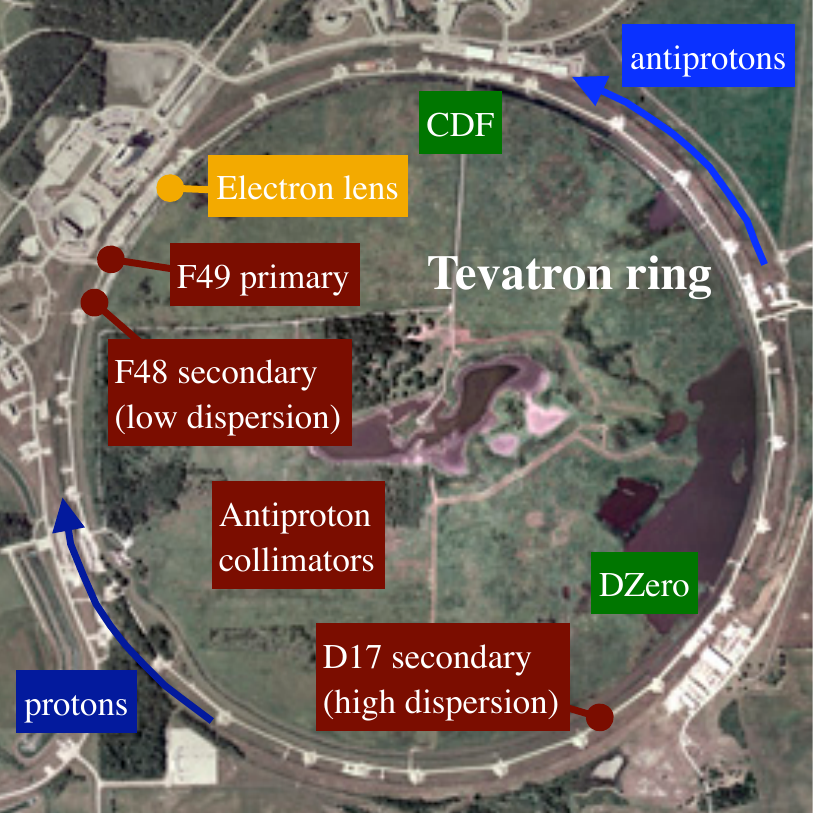} \hfill
\includegraphics[width=0.48\textwidth]{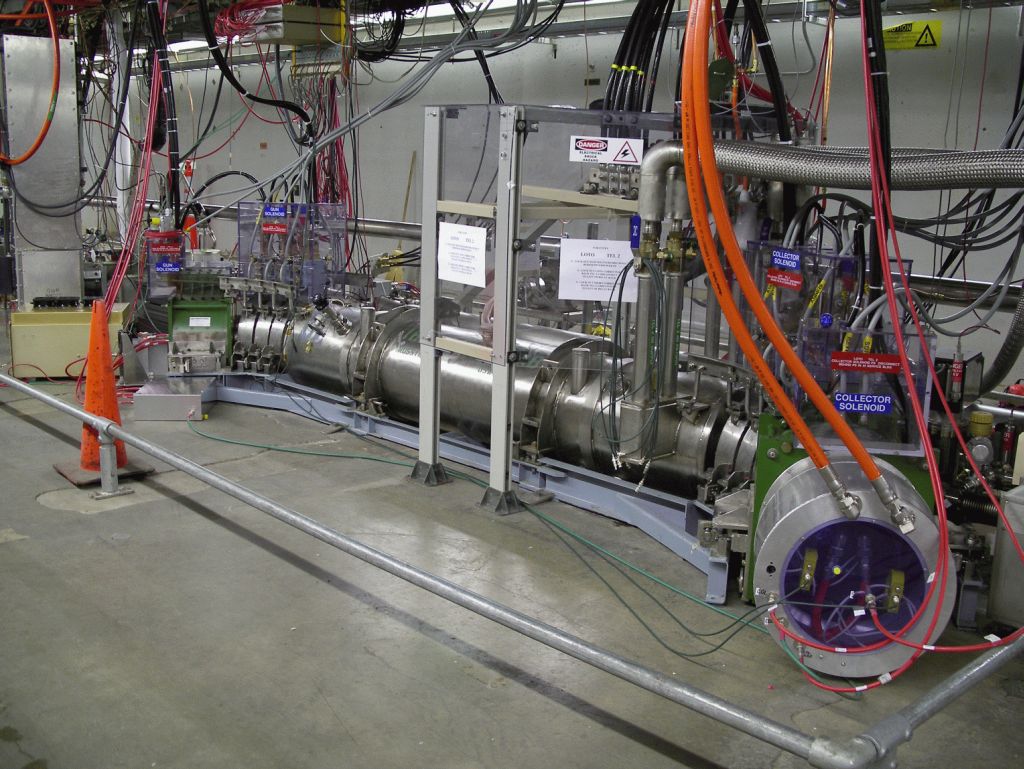}
\caption{(left)~Layout of the hollow-beam collimation
  experiments. (right)~Photograph taken in the Tevatron tunnel of the
  electron lens (TEL2) used as hollow-beam
  collimator.\label{fig:hebc.layout}}
\end{figure}

\begin{figure}[b]
\begin{tabular}{cc}
\includegraphics[height=0.4\textwidth]{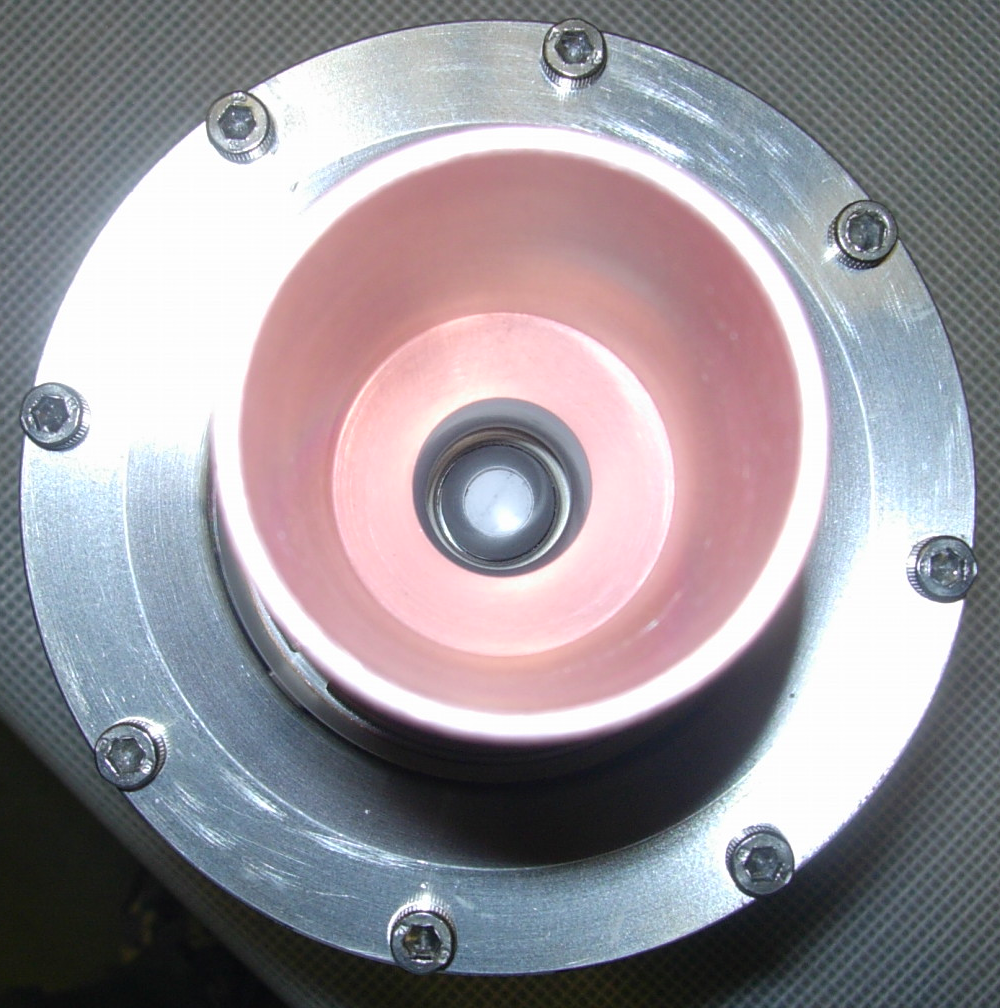} &
\includegraphics[height=0.4\textwidth]{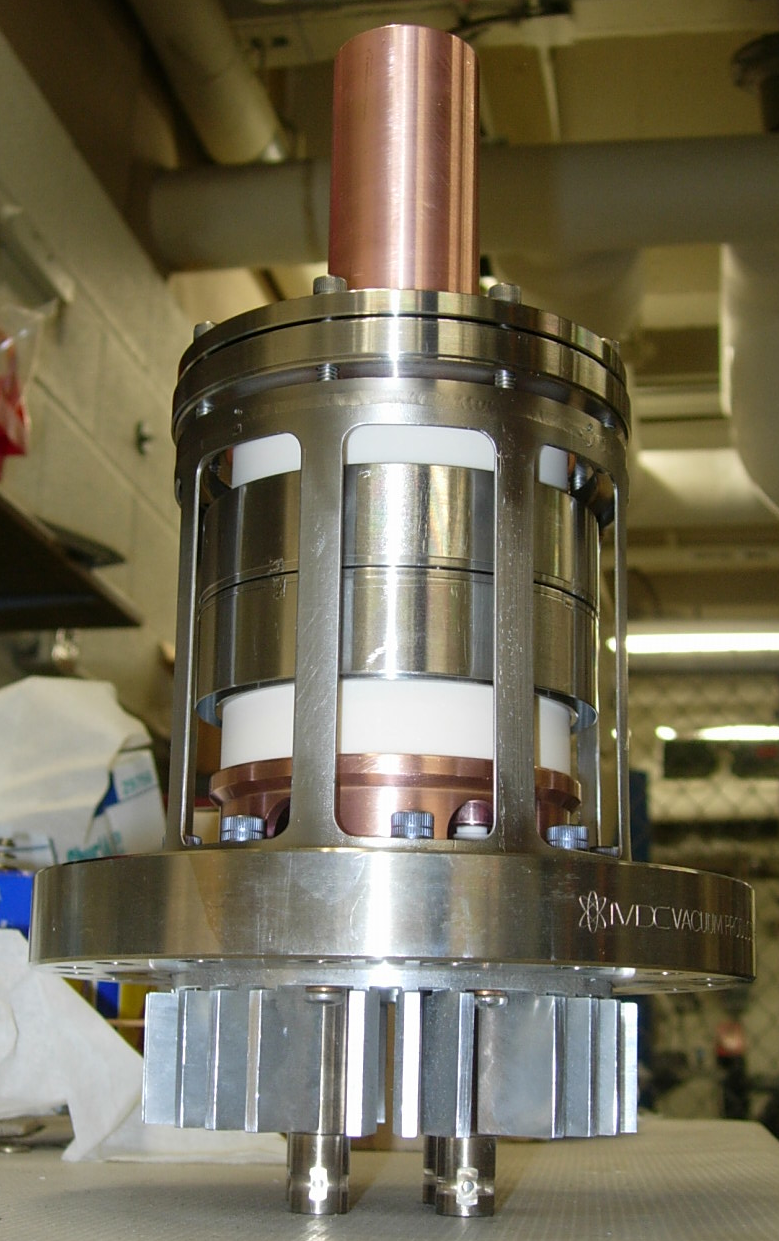} \\
\includegraphics[width=0.48\textwidth]{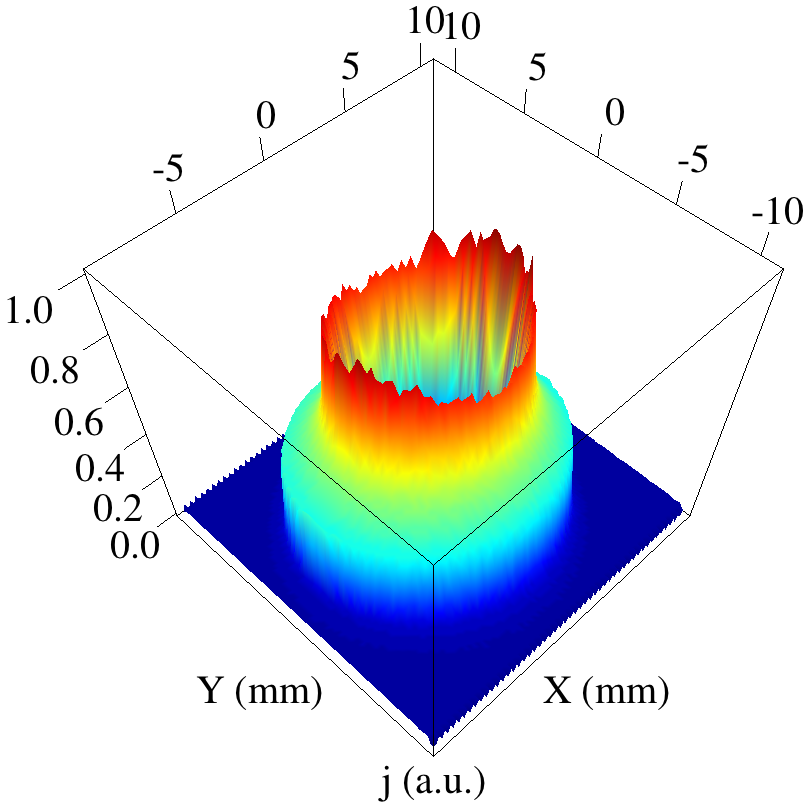} &
\includegraphics[width=0.48\textwidth]{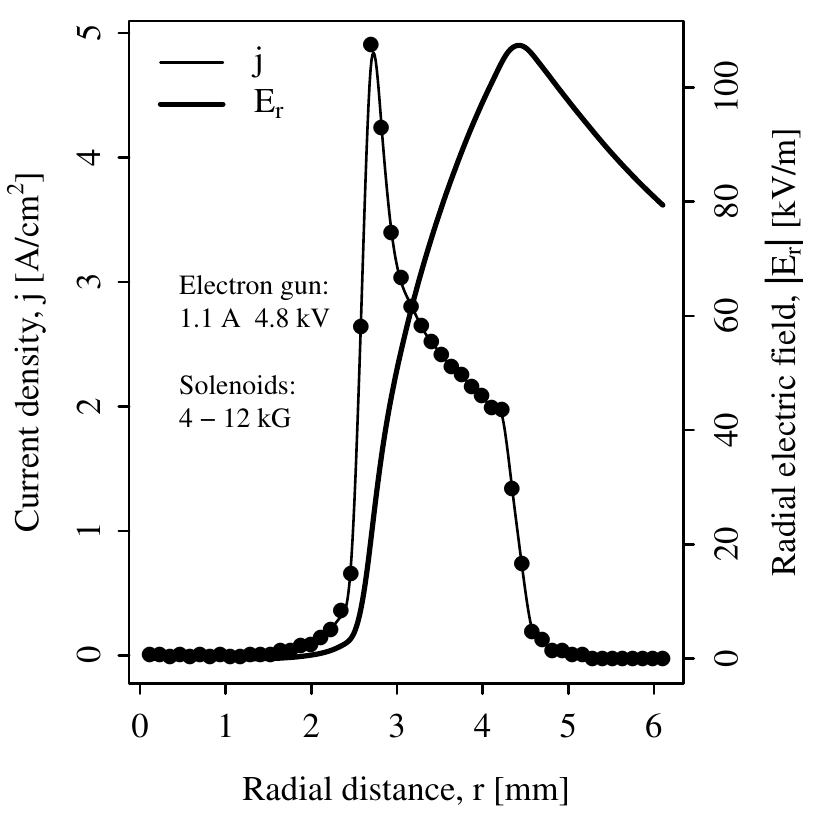} \\
\end{tabular}
\caption{15-mm hollow electron gun used in the Tevatron experiments:
  (top~left)~photograph of the copper anode and of the hollow tungsten
  cathode; (top~right)~photograph of the assembled gun;
  (bottom~left)~example of measured transverse current profile;
  (bottom~right)~example of radial current density distribution and
  electric field in the overlap region for a total current of 1.1~A
  and a magnetic compression factor of $\sqrt{(\q{4}{kG}) /
    (\q{12}{kG})} = 0.58$.\label{fig:egun}}
\end{figure}

\begin{figure}[b]
\includegraphics[width=0.8\textwidth]{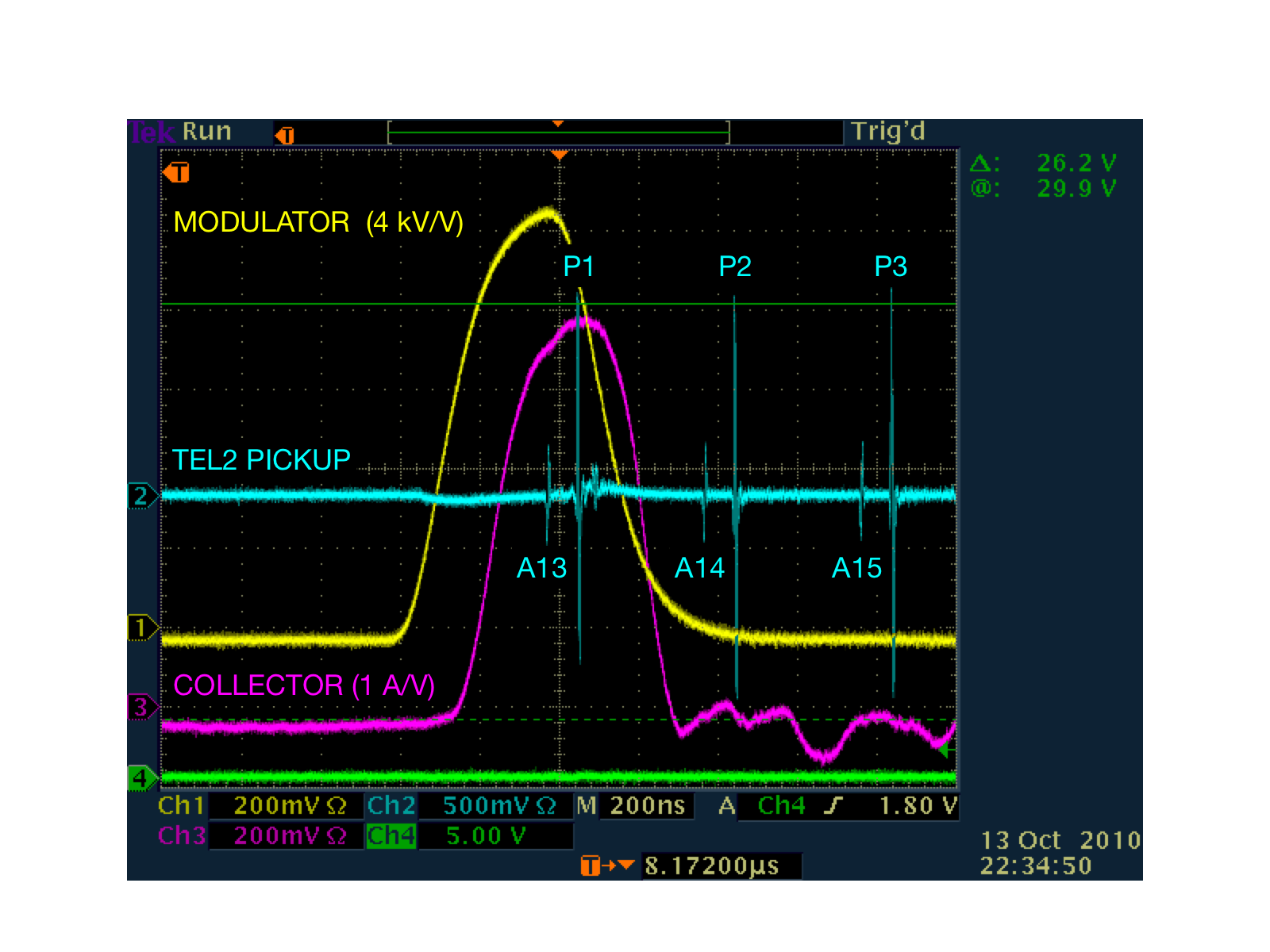}
\caption{Synchronization of the electron beam pulse with an antiproton
  bunch: modulator voltage (yellow); electron current in the collector
  (magenta); pickup signal (cyan) showing three proton bunches
  (P1--P3), three antiproton bunches (A13--A15), and the derivative of
  the electron pulse.\label{fig:sync}}
\end{figure}

\begin{figure}[b]
\includegraphics[width=0.48\textwidth]{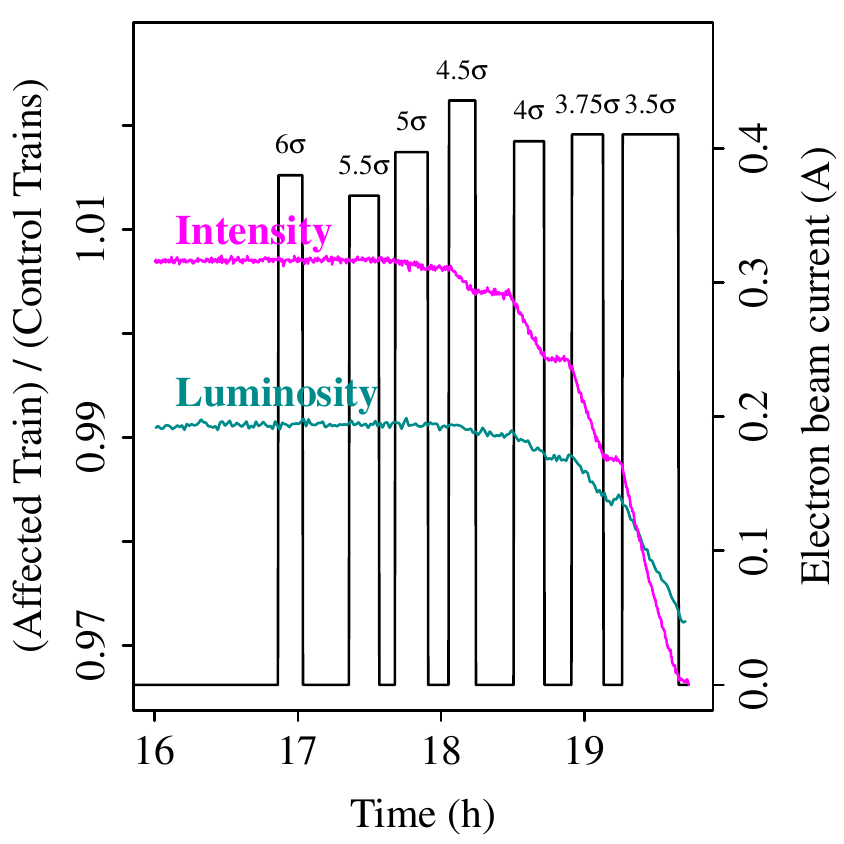} \hfill
\includegraphics[width=0.48\textwidth]{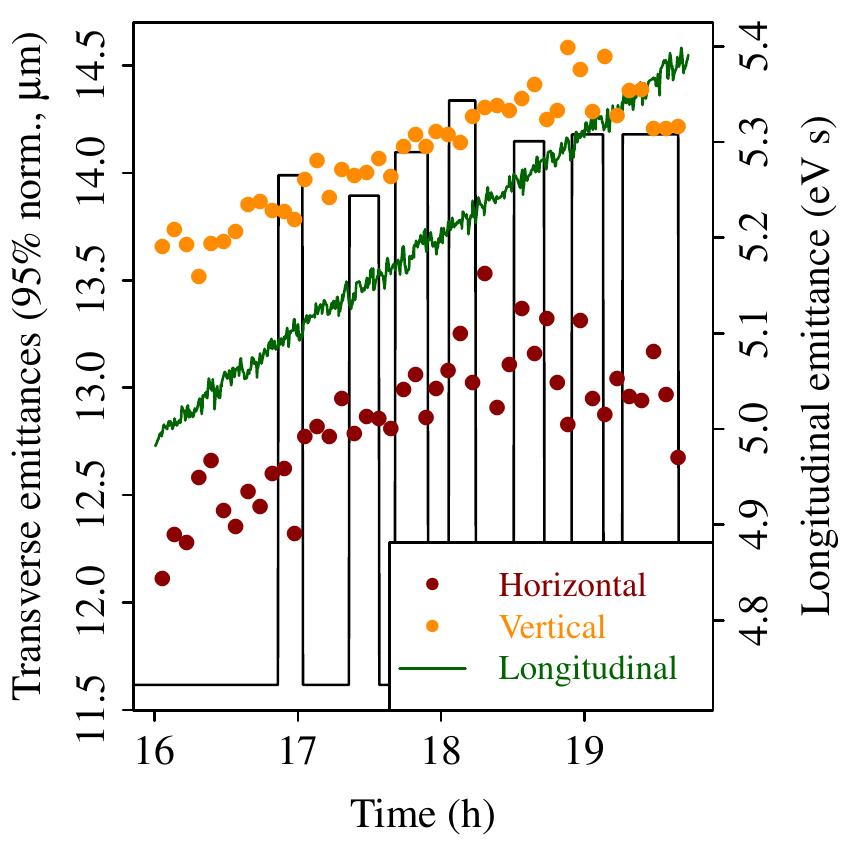}
\caption{Effects of the hollow electron beam collimator on 980-GeV
  antiprotons in the Tevatron at the end of a regular collider store,
  for different values of the hole radius. (left)~Relative intensity
  and luminosity of the affected bunch train; (right)~transverse and
  longitudinal emittances of the affected bunch train during the same
  experiment. The black trace represents the intensity of the electron
  beam.\label{fig:scraping}}
\end{figure}

\begin{figure}[b]
\includegraphics[width=0.48\textwidth]{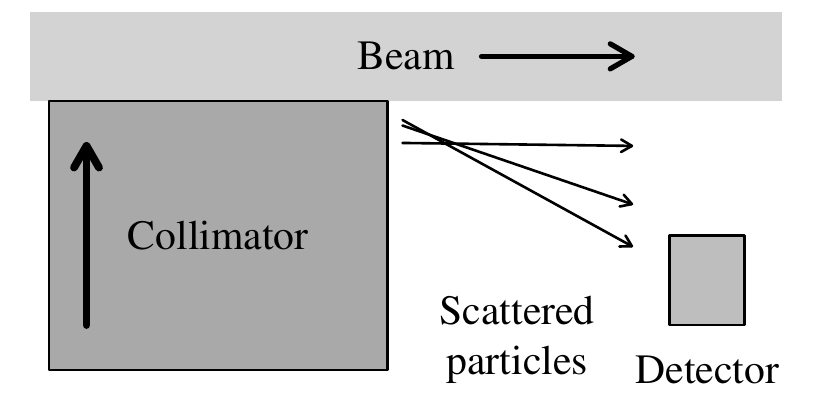} \hfill
\parbox[c]{0.48\textwidth}{\includegraphics[width=0.48\textwidth]{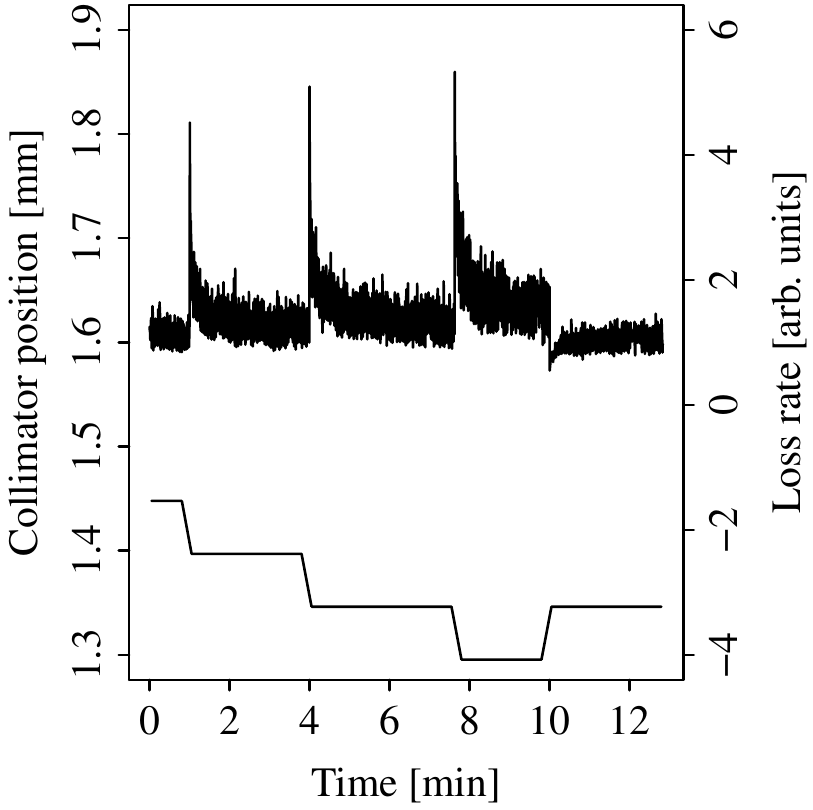}} \\
\includegraphics[width=0.48\textwidth]{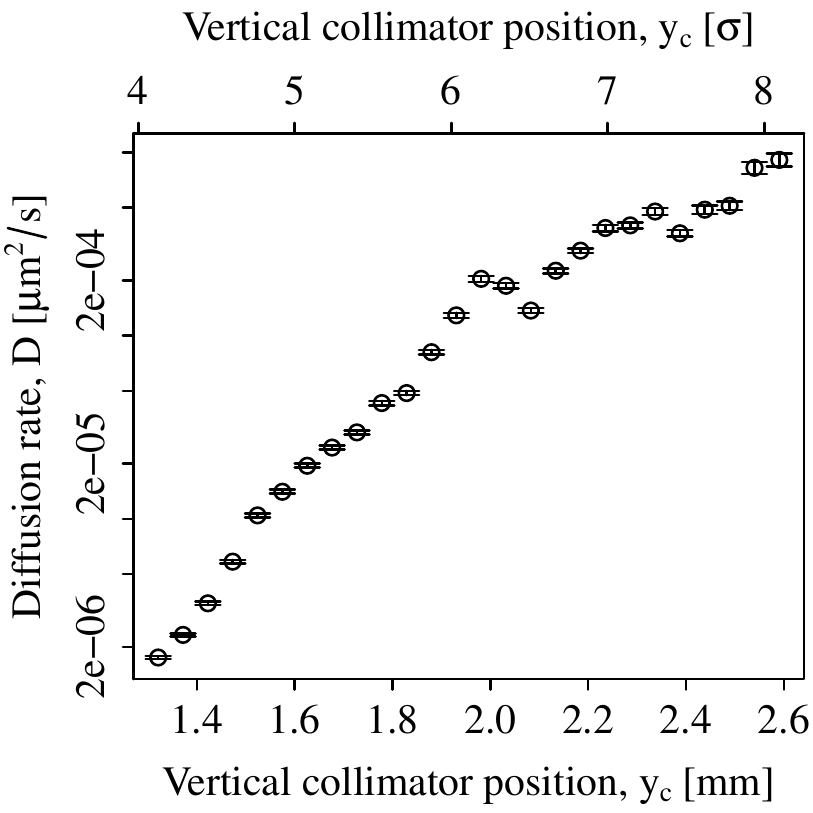} \hfill
\includegraphics[width=0.48\textwidth]{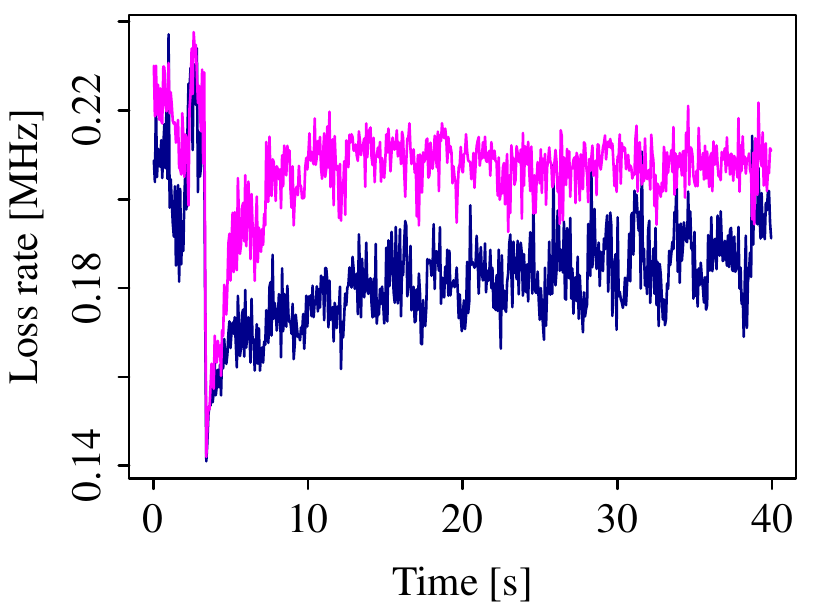}
\caption{Measurements of transverse beam diffusion with collimator
  scans. (top~left)~Schematic view of the experiment: local losses are
  measured as a collimator is moved inward or outward in small
  steps. (top~right)~Example of measured loss rates as the F48
  secondary antiproton collimator is moved inward (3~steps) and
  outward (1~step) by \q{50}{\mu m}. (bottom~left)~Measurement of
  diffusion rate vs.\ amplitude with a collimator scan in the
  Tevatron. (bottom~right)~Simultaneous measurement of local loss
  rates in response to an outward step for the control bunch train
  (blue) and the bunch train affected by the hollow electron beam
  (magenta), demonstrating diffusion enhancement.\label{fig:cscan}}
\end{figure}

\begin{figure}[bpt]
\includegraphics[width=0.48\textwidth]{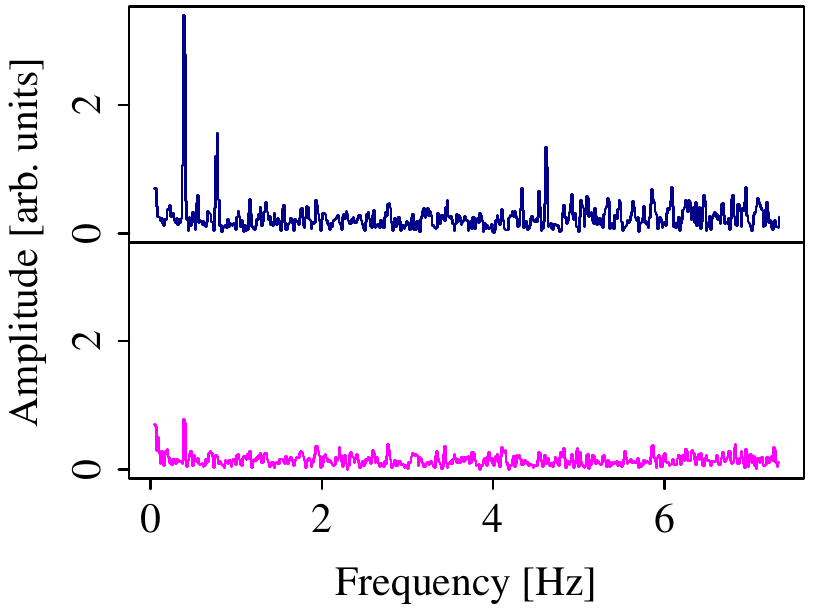} \hfill
\includegraphics[width=0.48\textwidth]{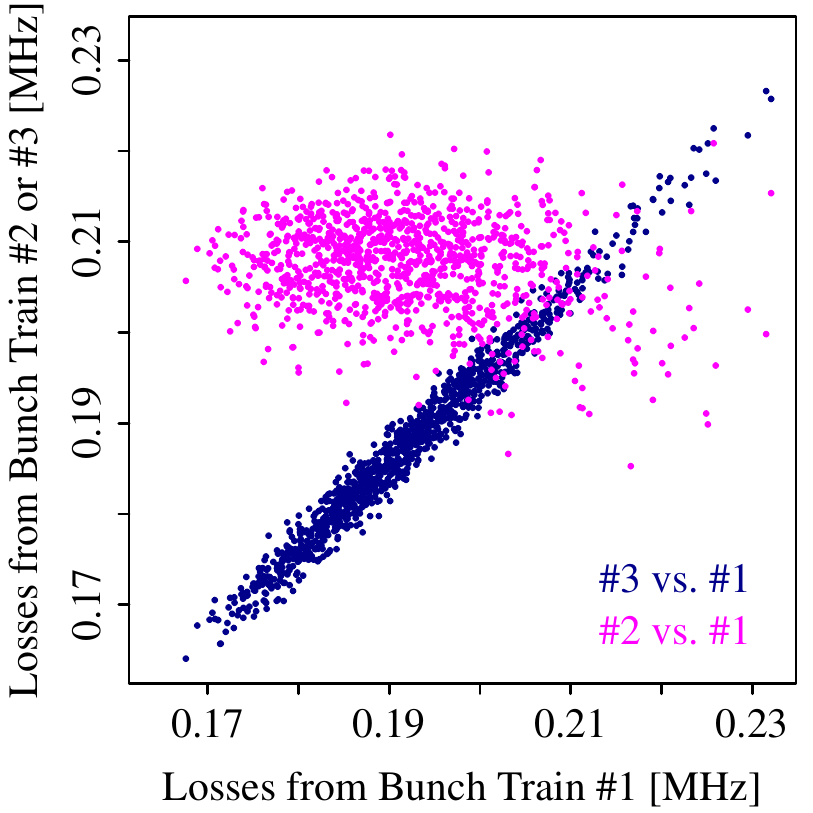}
\caption{Simultaneous measurement of steady-state loss rates for the
  two control bunch trains (\#1 and \#3) and for the bunch train
  affected by the hollow electron beam (\#2). (left)~Frequency
  spectrum of losses: control train \#1 (blue) and affected train \#2
  (magenta); coherent peaks due to vibrations are
  suppressed. (right)~Correlation between loss rates: \#3 vs. \#1
  (blue), showing that beam jitter dominates loss fluctuations; \#2
  vs. \#1 (magenta) shows that correlation is removed and peaks due to
  beam jitter are mitigated.\label{fig:jitter}}
\end{figure}

The concept of hollow electron beam collimation was tested
experimentally in the Fermilab Tevatron collider
(Figure~\ref{fig:hebc.layout}). In the Tevatron, 36 proton bunches
collided with 36 antiproton bunches at an energy of 980~GeV per
beam. Each particle species was arranged in 3~trains of 12~bunches
each. Initial beam intensities were typically \q{3\times
  10^{11}}{protons/bunch} and \q{10^{11}}{antiprotons/bunch}. Beam
lifetimes ranged between 10~h and 100~h. There were 2 head-on
interaction points, corresponding to the CDF and the DZero
experiments. The maximum luminosity was \q{4\times 10^{32}}{cm^{-2} \,
  s^{-1}}.  The machine operated with betatron tunes near~20.58.

A 15-mm-diameter hollow electron gun was designed and built
(Figure~\ref{fig:egun}). It was based on a tungsten dispenser cathode
with a 9-mm-diameter hole bored through the axis of its convex
surface. The peak current delivered by this gun was 1.1~A at 5~kV. The
current density profile was measured on a test stand by recording the
current through a pinhole in the collector while changing the position
of the beam in small steps. A sample measurement is shown in
Figure~\ref{fig:egun}. In August~2010, the gun was installed in one of
the Tevatron electron lenses (TEL2), located in the A~sector of the
ring. Experiments were done with antiprotons because of their smaller
size and because of the configuration of the Tevatron collimation
system. For antiprotons, tungsten primaries (F49 in
Figure~\ref{fig:hebc.layout}) were located downstream of the electron
lens. Stainless-steel secondaries were placed at the F48 and D17
locations in the ring.

In the electron lens, protons and antiprotons were separated
transversely and in time. The transverse separation was about 6~mm
both horizontally and vertically (Figure~\ref{fig:hebc.concept}). The
radius of the hole was controlled by the ratio of solenoid
fields. Three corrector coils were used to align the electron beam
with the circulating beam. Thanks to the special high-voltage
modulator, the electron beam could be synchronized practically with
any group of bunches. In the oscilloscope picture of
Figure~\ref{fig:sync}, one can see the short proton and antiproton
bunches in cyan, and the shortest possible electron pulse in
magenta. Alignment of the beams was found to be reliable and
reproducible. The stability of the system of three beams was not an
issue at solenoid fields above 10~kG.

Experiments in the Tevatron started in October~2010 and ended with the
shutdown of the machine in September~2011. Many observables such as
particle removal rates, effects on the core, diffusion enhancement,
collimation efficiency and loss rate fluctuations were measured as a
function of electron lens parameters: beam current, relative
alignment, hole radius, pulsing pattern, and collimator
configuration. Preliminary results were presented in
Refs.~\cite{Stancari:PAC:2011, Stancari:PRL:2011,
  Stancari:IPAC:2011}. Here, a few examples of the main effects are
shown.

\subsection{Particle removal rates}

The first question we addressed was the particle removal rate. Can
these small transverse kicks have a detectable effect?  The experiment
in Figure~\ref{fig:scraping} (left) shows the scraping effect. The
electron lens was aligned and synchronized with the second antiproton
bunch train, and then turned on and off several times at the end of a
collider store. The electron beam current was about 0.4~A and the
radius of the hole was varied between 6 and 3.5 standard deviations
($\sigma$) of the vertical beam size. The black trace is the
electron-lens current. To isolate the effect of the hollow beam,
Figure~\ref{fig:scraping} (left) shows the ratio between the intensity
of the affected train and the average intensity of the other two
control trains. One can clearly see the smooth scraping effect. The
corresponding removal rates are a few percent per hour, and increase
with decreasing hole radii. One can also see that there were no
adverse effects with large hole sizes. This allowed us to do many
experiments parasitically during regular collider stores.

\subsection{Effects on the beam core}

Which particles are being removed?  Whether there are any adverse
effects on the core of the circulating beam is a concern. The overlap
region is not a perfect hollow cylinder, due to asymmetries in gun
emission, to evolution under space charge of the hollow profile, and
to the bends in the transport system.

We estimated the effects on the core from several points of view:
particle removal rates, emittance growth, luminosity, and diffusion
enhancement vs.\ amplitude. First, one can see from
Figure~\ref{fig:scraping} (left) that no decrease in intensity was
observed with large hole sizes, when the hollow beam was shadowed by
the primary collimators. This implies that the circulating beam was
not significantly affected by the hollow electron beam surrounding it,
and that the effect on beam intensity of residual fields near the axis
was negligible.

Secondly, one can observe the evolution of the emittances. In
Figure~\ref{fig:scraping} (right), the average emittances of the
affected bunch train are shown. If there was emittance growth produced
by the electron beam, it was much smaller than that driven by the
other two main factors, intrabeam scattering and beam-beam
interactions. As expected, for small hole sizes, suppression of the
beam tails translated into a reduction in measured transverse
emittances.

The effect of halo removal can also be observed by comparing beam
scraping with the corresponding decrease in luminosity. Luminosity is
proportional to the product of antiproton and proton populations, and
inversely proportional to the overlap area. If antiprotons were
removed uniformly and the other factors were left unchanged,
luminosity would decrease by the same relative amount. If the hollow
beam caused emittance growth or proton loss, luminosity would decrease
even more. Because of the smaller relative contribution to luminosity
of halo particles, a smaller relative change in luminosity is a clear
indication of halo scraping. In Figure~\ref{fig:scraping} (left), one
can see how the luminosity for the affected bunch train changed with
time relative to the average luminosity of the control bunch
trains. The dark cyan trace is the ratio between affected and control
trains. The corresponding relative luminosity decay rates are smaller
that those of intensity. As expected, luminosity decay rates approach
intensity decay rates as the radius of the hole is decreased.

\subsection{Diffusion enhancement}

It is possible to measure transverse beam diffusion rates as a
function of betatron amplitude by observing the time evolution of
losses as collimators are moved in small steps~\cite{Mess:NIM:1994,
  Seidel:PhD:1994, Stancari:arXiv:2011.diff, Stancari:IPAC:2011b}
(Figure~\ref{fig:cscan}, top~left). The main features of the response
of local losses to small inward collimator steps are a sharp peak and
a decay proportional to the inverse square root of time. From this
decay the diffusion rate at the location of the collimator can be
extracted. Figure~\ref{fig:cscan} (top right) shows an example of a
collimator scan. The vertical F48 secondary collimator was moved in
50-$\mu$m steps towards the beam center. All other collimators were
retracted. The characteristic decay time is a function of collimator
position. These loss spikes are what constrained the tightest
collimator settings and limited their setup. Even when the collimators
were moved in micron-size steps, below about 4$\sigma$ in transverse
beam size these spikes could exceed the quench limits in the Tevatron
superconducting magnets. Figure~\ref{fig:cscan} (bottom left) shows an
example of diffusion measurements vs.\ amplitude at the end of a
regular collider store. Similar to the results obtained in the HERA
machine at DESY~\cite{Mess:NIM:1994, Seidel:PhD:1994}, a strong
dependence of the diffusion coefficient with amplitude is
observed. More details can be found in
Refs.~\cite{Stancari:arXiv:2011.diff, Stancari:IPAC:2011b}.

We are interested in how the hollow beam affects diffusion. New
scintillator paddles were installed near one of the antiproton
secondary collimators during a short shutdown in March~2011. These
loss monitors were gated to individual bunch trains. With this device
we could measure diffusion rates, collimation efficiencies and loss
spikes simultaneously for the affected and the control bunch
trains. An example of what can be observed by comparing losses from
different bunches is shown in Figure~\ref{fig:cscan} (bottom
right). The electron lens was aligned and synchronized with bunch
train~\#2 with a peak current of 0.9~A.  The plot shows the response
of local losses to an outward collimator step. In this case, one
observes a dip instead of a spike. The difference in diffusion times
between the control and the affected train is apparent. It corresponds
to an enhancement of the diffusion rate by a factor~10. Because the
steady-state loss rate is the product of diffusion coefficient and
gradient of the beam population, the fact that it is only 10\% higher
for the affected train indicates that the halo population was reduced
by approximately a factor~10 as well.

\subsection{Mitigation of beam jitter}

In Figure~\ref{fig:cscan} (bottom right), one may notice some of the
periodic losses due to beam jitter in the control train. They are
suppressed in the affected train: another indication of halo removal.
This effect is also apparent in the Fourier spectrum of gated losses
(Figure~\ref{fig:jitter}, left). Normally, the spectra showed peaks
corresponding to mechanical vibrations caused, for instance, by the
Main Injector acceleration ramp (0.3~Hz) or by the compressors of the
Central Helium Liquefier (4.6~Hz). The electron beam acting on the
second bunch train suppressed these periodic losses. This is another
manifestation of the reduction of tails.

A better understanding of the distribution of these losses comes from
the analysis of their correlations. The blue points in
Figure~\ref{fig:jitter} (right) show the losses coming from train~\#3
vs.\ those from train~\#1. These are the two control bunches. One can
see random fluctuations of the order of 2~kHz out of 0.2~MHz. But the
main effect is a very high correlation. Most of the fluctuations are
not random, but coherent. They can be attributed to beam jitter. The
effect of the hollow beam is to eliminate this correlation: average
losses are increased, instantaneous losses are randomized, and the
spikes are reduced. This effect can be interpreted as an increase in
diffusion and a decrease in tail population, which translates into a
reduced sensitivity to beam jitter.

\section{Conclusions and outlook}

Thanks to the dedication of all the people involved in the project,
the hollow electron beam collimator now appears to be a viable tool
for high-power circular accelerators. Many experimental observations
of the scraping effects were made at the Tevatron using the existing
electron lenses equipped with a hollow electron gun designed for this
purpose.

A new 25-mm gun was designed and built. It will be tested at the
Fermilab electron-lens test stand to investigate the technical
feasibility of larger currents (about 3~A at 5~kV) and beam radii.  We
also plan to increase our effort in simulations to understand
hollow-beam collimation in the Tevatron and extend it to other
machines.

Some of the electron lens hardware will become available after the
Tevatron shutdown. Recently, both the DOE LARP Review
Committee~\cite{DOE.LARP.Review:2011} and the LHC Collimation
Review~Committee~\cite{LHC.Collimation.Review:2011} recommended
continuing the experimental program at CERN in the SPS or in the
LHC. In collaboration with the LHC Collimation Working Group, we are
studying whether a hollow-beam scraper is applicable to the CERN
machines.

\begin{acknowledgments}
  The work on hollow electron beam collimation was done in
  collaboration with A.~Valishev, J.~Annala, T.~Johnson, G.~Saewert,
  V.~Shiltsev, D.~Still, and L.~Vorobiev of Fermilab.

  This project would not have been possible without the support of the
  Fermilab Accelerator Division and of the CDF and DZero
  collaborations. In particular, I would like to thank M.~Convery,
  C.~Gattuso, and R.~Moore for their invaluable support. Fermi
  Research Alliance, LLC operates Fermilab under Contract
  No.~DE-AC02-07CH11359 with the US Department of Energy. This work
  was partially supported by the US LHC Accelerator Research Program
  (LARP).
\end{acknowledgments}

\end{document}